\documentclass[prl,twocolumn]{revtex4}
\usepackage{graphicx}
\usepackage{hyperref}
\usepackage[section]{placeins}
\usepackage{amsmath}
\usepackage{mathtools}
\usepackage{amssymb}
\usepackage{dsfont}
\usepackage{bbold}
\usepackage{ulem}
\newcommand{\ket}[1]{|#1\rangle}             
\newcommand{\bra}[1]{\langle#1|}             
\newcommand{\braket}[2]{\langle#1|#2\rangle} 

\begin{document}

\title{Reconstructing high-dimensional two-photon entangled states via compressive sensing}

\author{Francesco Tonolini$^{1}$, Susan Chan$^{1}$, Megan Agnew$^{1}$, Alan Lindsay$^{1, 2}$, Jonathan Leach$^{1}$}

\affiliation{$^1$SUPA, School of Engineering and Physical Sciences, Heriot-Watt University, Edinburgh EH14 4AS, UK}\affiliation{$^2$Applied and Computational Mathematics and Statistics, University of Notre Dame, USA} 

\date{\today}

\begin{abstract}
Accurately establishing the state of large-scale quantum systems is an important tool in quantum information science; however, the large number of unknown parameters hinders the rapid characterisation of such states, and reconstruction procedures can become prohibitively time-consuming.  Compressive sensing, a procedure for solving inverse problems by incorporating prior knowledge about the form of the solution, provides an attractive alternative to the problem of high-dimensional quantum state characterisation.  Using a modified version of compressive sensing that incorporates the principles of singular value thresholding, we reconstruct the density matrix of a high-dimensional two-photon entangled system.   The dimension of each photon is equal to $d= 17$, corresponding to a system of 83521 unknown real parameters.  Accurate reconstruction is achieved with approximately 2500 measurements, only 3\% of the total number of unknown parameters in the state.  The algorithm we develop is fast, computationally inexpensive, and applicable to a wide range of quantum states, thus demonstrating compressive sensing as an effective technique for measuring the state of large-scale quantum systems.
\end{abstract}

\maketitle

{\large \noindent{\bf{Introduction}}}

\noindent Many areas of quantum mechanics require the efficient and accurate measurement of entangled states. Perhaps the most traditional and widely adopted way of doing so is through full tomographic reconstruction \cite{DAriano2003}, a technique that performs a series of independent measurements on the system in order to uniquely identify its nature. However, the complexity of such a method dramatically increases with increasing dimension of the system, and fully measuring the state of two entangled objects, each of $d$ dimensions, requires at least $d^4$ measurements \cite{James2001}. As a result, full tomographic reconstruction is effective only at low dimensions and is otherwise prohibitively time consuming and computationally expensive.

Large-dimensional states are necessary for quantum computation and for certain quantum information protocols.  Monz {\it et al.} reported the generation of a 14-qubit entangled state using trapped ions \cite{Monz2011}, and Yao {\it et al.} reported the generation of an 8-photon entangled state \cite{Yao2012}, although neither reported the density matrix for their respective states.   Zhang {\it et al.} performed quantum tomography of a hybrid optical detector with over a million free parameters \cite{Zhang2012}.  However, to date, the largest density matrix reported for an entangled state is that of H\"{a}ffner {\it et al.}, who recorded the density matrix of 8 trapped ions \cite{Haffner2005}.

Compressive sensing, which originates from the field of signal processing, provides a very efficient mechanism to establish properties of an unknown system with limited observations (see, e.g., Cand\`es \cite{Candes2006} and references therein).   Compressive sensing uses prior assumptions in order to reduce the number of possible solutions, which can drastically reduce both measurement and processing time.  Consequently, it is possible to establish descriptions of very large systems that could previously not be explored.  This principle is used extensively in the fields of image reconstruction \cite{Baraniuk2008} and medical tomography \cite{Lustig2007}, and it has recently been adopted in various areas of quantum science \cite{Gross2010,Howland2011,Shabani2011,Howland2013,Howland2014,Howland2014b,Mirhosseini2014}.

In this paper we propose and outline a compressive sensing technique that is able to successfully reconstruct the density matrix of near-pure entangled states of high dimensions. We implement this method to reconstruct a pure state of two 17-dimensional photons entangled in their orbital angular momentum. The recovery of the state is achieved by employing only 3\% of the measurements that full tomographic reconstruction would require. The full procedure, including measuring and post-processing, takes approximately three hours on a standard desktop computer. Our data processing algorithm is similar to the singular value thresholding algorithm detailed in \cite{Cai2010}; however, its design is specifically adapted for near-pure entangled state reconstruction. The procedure is fast, computationally inexpensive, and robust to noise.

\vspace{.5 cm}
{\large \noindent{\bf{Results}}}

{\noindent{{\bf{Theoretical description of compressive sensing and quantum tomography.}}}} 

\noindent Compressive sensing is a data-processing technique widely used in different signal reconstruction applications. Its aim is to find the solution to underdetermined linear systems, under the assumption that such a solution is sparse in some basis. Such problems can be posed in the following way:
\begin{align}\label{stcompress}
\min\|f(\textbf{x})\|_1 \quad s.t. \quad  A\textbf{x}=\textbf{b},
\end{align}
where $\textbf{x} \in \mathbb{C}^{N \times 1}$ represents a vector describing the measured object; $A \in \mathbb{C}^{M \times N}$ is the matrix of measurements, with $M \ll N$; $\textbf{b} \in \mathbb{C}^{M \times 1}$ is the vector of measurement results; $\|\cdot\|_1$ denotes the $\ell 1$ norm of the vector; and $f$ is a transformation to a space in which $f(\textbf{x})$ has a sparse representation.

In the specific case of quantum state tomography, the aim is to reconstruct an unknown near-pure density matrix, using an under-sampled set of measurements, under the assumption that such a matrix is low rank. The problem to be solved is then \cite{Gross2010,Liu2012}
\begin{align}\label{stcompress2}
\min{ \|\hat{\rho}\|_{\rm Tr} }  \quad s.t. \quad  A\vec{\rho}=\vec{p}, \quad  {\rm Tr}(\hat{\rho}) = 1, \quad \hat{\rho} = \hat{\rho}^{\dagger}.
\end{align}
Here, $\hat{\rho}$ is the density matrix to be reconstructed, while $\vec{\rho} \in \mathbb{C}^{N \times 1}$ is the density matrix in vector form; $A \in \mathbb{C}^{M \times N}$ is the matrix of measurements; $\vec{p} \in \mathbb{C}^{M \times 1}$ is the vector of resulting probabilities; and $\|\cdot\|_{\rm Tr}$ stands for the trace norm of the matrix. The rows of the measurement matrix $A$ are individual measurement vectors $A_i$, and the elements of the vector $\vec{p}$ are the corresponding probabilities $p_i$.

\vspace{.5 cm}
\noindent{\bf{The algorithm.}}

\noindent
We develop an operation-projection method similar to the singular value thresholding algorithm shown in \cite{Cai2010} and implemented in \cite{Gross2010}; however, we significantly modify its design in order to make use of the known features of near-pure entangled states. 
The algorithm requires an initial guess matrix to begin the procedure. The protocol then has two main stages: (i) the operations on the current matrix $\hat{\rho}$ to impose the desired characteristics and (ii) the projection of the resulting answer in vector form $\vec{\rho}$  onto the solution space. Applying these steps repeatedly constitutes an iterative procedure to approach the target solution. We interchange between the matrix form and vector form when implementing the operation and projection stages respectively.

\begin{figure*} [ht]
\includegraphics[]{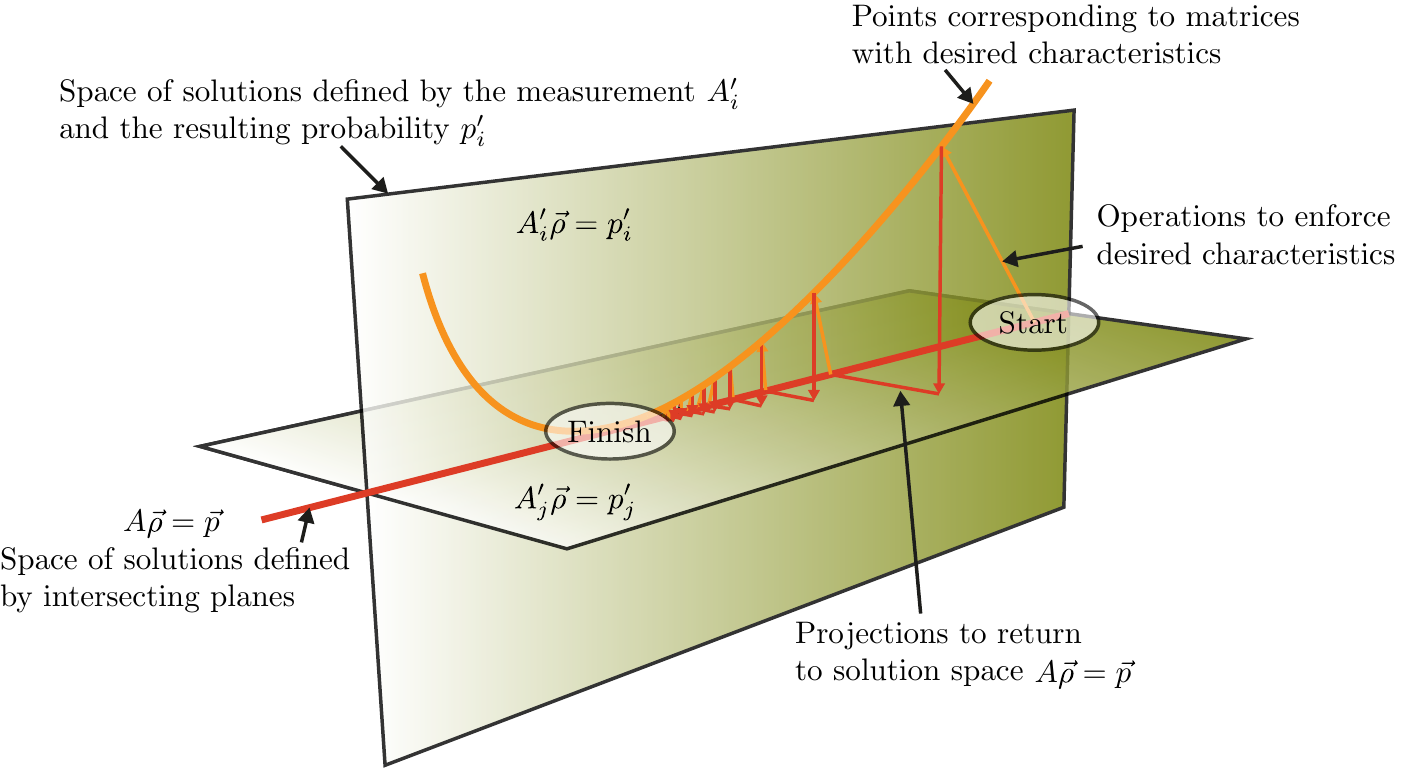}
 \caption{\footnotesize{{ A schematic representation of the compressive sensing problem. The two green planes $A^{\prime}_{i}\vec{\rho}=p^{\prime}_{i}$ and $A^{\prime}_{j}\vec{\rho}=p^{\prime}_{j}$ each correspond to individual measurements and represent two different solution spaces. The intersection of the two planes corresponds to the set of all solutions belonging to the combined space $A\vec{\rho}=\vec{p}$, indicated by the red line. The curved line represents a set of potential solutions in the space that retain the desired characteristics of our answer. Our algorithm works by iterating between the set of solutions with the desired characteristics (orange line) and the set of solutions belonging to $A\vec{\rho}=\vec{p}$ (red line).  After a number of iterations, the 
 algorithm converges to the solution of $A\vec{\rho}=\vec{p}$ that possesses the desired characteristics.}}}
 \label{coolfigure}
\end{figure*}

\begin{figure*} 
\includegraphics[]{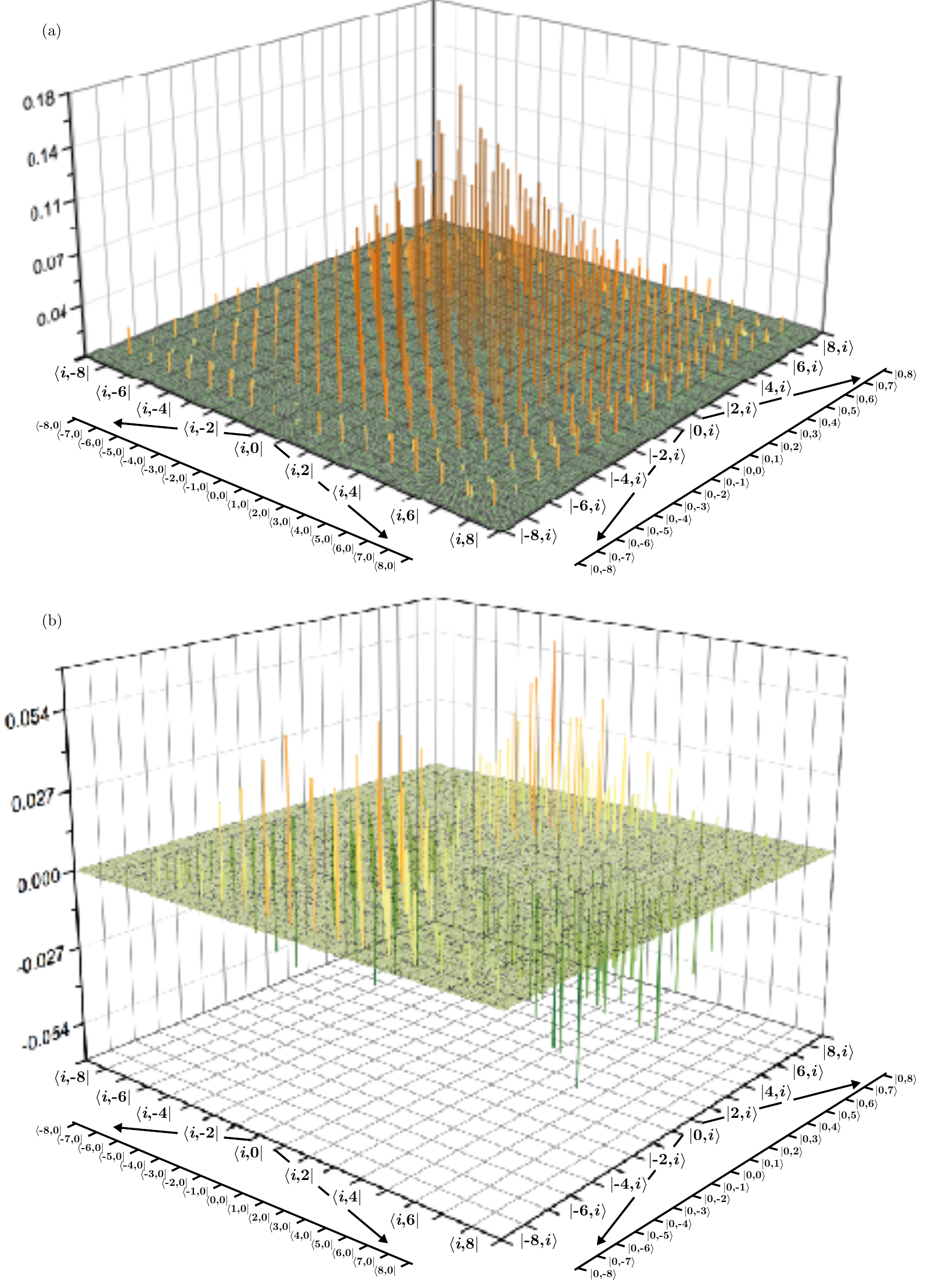}
 \caption{\footnotesize{{(a) The real part of the recovered density matrix.  The dimension of each photon is equal to $d$ = 17, so that the total dimension of the combined space is equal to 83521.   The index $i$ runs from $i=-8$ to $8$. (b) The imaginary part of the recovered density matrix.}}}
 \label{real}
\end{figure*}

In the operations stage, two steps are performed:  First, the rank of the matrix is reduced by thresholding the eigenvalues below a certain level.   This is achieved by decomposing the matrix into its eigenvalues and eigenvectors, setting the eigenvalues below the chosen threshold to zero, and then recomposing the matrix using
\begin{align}\label{stcompress3}
\hat{\rho} = \sum \lambda_i\ket{\varphi_i}\bra{\varphi_i},
\end{align}
where $\lambda_i$ is the $i^{th}$ eigenvalue and $\varphi_i$ the corresponding eigenvector. Second, to make use of the known sparsity characteristic and trace properties of the solution, we apply a thresholding operation on the individual matrix elements and normalise the result to have trace equal to unity.  We achieve this by setting the elements that have modulus smaller than a chosen value to zero and dividing the matrix by its trace. 

After the operation, the resultant matrix $\hat{\rho}_0$ has the desired characteristics of the solution; however, it no longer belongs to the linear space defined by the measurements $A$ and probabilities $\vec{p}$.  To return the matrix $\hat{\rho}_0$ to the space defined by $A\vec{\rho}=\vec{p}$, we then implement the projection stage of the procedure.  In order to describe the projection stage, we first introduce a geometrical formalism of the problem.

Each measurement vector $A_i$ and corresponding probability $p_i$ represents a hyperplane in a space of $N$ dimensions, where $N$ is the number of elements in $\vec{\rho}$. The intersection of these hyperplanes represents the set of all solutions to the system $A\vec{\rho}=\vec{p}$. A simplified version of this concept is shown in Fig.~\ref{coolfigure}, where two intersecting hyperplanes are represented as two-dimensional planes, and their intersection as a line.   

After the operations stage, the matrix $\hat{\rho}_0$ is reshaped into vector form $\vec{\rho}_0$ so that it can be projected onto the intersection of the hyperplanes defined by the linear system. The projection procedure is simple and computationally inexpensive if the hyperplanes are all perpendicular to each other.  

However, although the matrix of random measurements $A$ is nearly orthonormal, there is small non-zero overlap between any two measurements $A_i$ and $A_j$ ($i \neq j$).  This is due to the physical limitations of the measurement procedure.  For this reason, we transform the system $A\vec{\rho} = \vec{p}$ into a new system $A^{\prime}\vec{\rho} = \vec{p}\,^{\prime}$, where $A^\prime$ is an orthogonal matrix. This is achieved by multiplying both $A$ and $p$ by a matrix $B$ such that $BA=A^\prime$. It is important to note that the system $A^{\prime}\vec{\rho} = \vec{p}\,^{\prime}$ is a mathematical construct and no longer directly relates to the measurements and their corresponding probabilities; however, the set of solutions it defines is exactly the same as that of the original system.

In order to obtain a solution $\vec{\rho}_s$ from the initial point $\vec{\rho}_0$, we progressively project $\vec{\rho}_0$ on each hyperplane in turn. This procedure is initiated by projecting the initial point $\vec{\rho}_0$ onto the first hyperplane, given by $A^{\prime}_{1}\vec{\rho} = p^{\prime}_{1}$, to find a new point $\vec{\rho}_1$. This new point is then projected onto the second hyperplane, and we continue in this fashion until the desired solution $\vec{\rho}_s$ is found. This occurs after $M$ projections, where $M$ is the number of measurements. Details of this projection procedure can be found in the Supplementary Materials.

Applying this operation-projection procedure repeatedly constitutes an iterative method that provides a solution exhibiting the desired characteristics and belongs to the linear system $A\vec{\rho}=\vec{p}$. The schematic outline of the algorithm is shown in Fig.~\ref{coolfigure}, where the orange and red arrows represent the operation and projection steps, respectively. The method is considered complete when the distance between consecutive iterates is below a predetermined tolerance.

\vspace{.5 cm}
\noindent{\bf{Noise correction.}}

\noindent
In our system, noise manifests itself as errors in the measured probabilities.  Such noise is unavoidable, and consequently, the density matrix that we recover $\vec{\rho}_r$ will not correspond to the desired solution to the problem $A\vec{\rho}=\vec{p}$; instead, it will be a solution to the system $A\vec{\rho}=\vec{p}+\Delta \vec{p}$, where $\Delta \vec{p}$ is a vector of errors on the true probabilities. This error in probabilities results in a solution $\vec{\rho}_r$ that is in fact some distance $\Delta \vec{\rho}$ from the desired solution $\vec{\rho}_d$ in the space in which the algorithm operates. There are many methods for finding the solution in the presence of error \cite{Beck2009,Cai2010}. In our case, we determine $\Delta \vec{\rho}$ and subtract it from $\vec{\rho}_r$. This corresponds to the operation
\begin{equation}
\vec{\rho}_d=\vec{\rho}_r-\Delta \vec{\rho}.
\end{equation}
We use \textit{a priori} knowledge of the desired state's characteristics to find $\Delta \vec{\rho}$ and systematically correct for noise in the system. Further details of our method can be found in the Supplementary Information.

\vspace{.5 cm}
 \noindent{\bf{Experimental implementation. }}

\noindent
We have performed an experimental recovery of the density matrix of a 17-dimensional two-photon state in the orbital angular momentum (OAM) degree of freedom, produced by parametric downconversion (see Methods for details). The dimension of each photon is equal to $d= 17$, so the dimension of the entire state is 83521.  The reconstruction is performed after 2506 projective measurements, which corresponds to only 3\% of the total number of unknown parameters in the state. The reconstruction of the state is shown in Fig.~\ref{real}.

The state that we measure exhibits strong anti-correlations in the OAM degree of freedom; that is to say that the OAM state $\ket{\ell}_S$ in the signal photon is correlated with the state $\ket{-\ell}_I$ in the idler photon. Additionally, the existence of the non-zero off-diagonal elements in the density matrix indicates a high degree of purity in the obtained state.  These two features combine to suggest a high degree of entanglement of the OAM modes.

To characterise the entanglement, we use the fidelity of the reconstructed state $\rho$ with the ideal, maximally entangled pure state
\begin{align}
\ket{\Phi} =\sum_{\ell} \frac{1}{d^2} \ket{-\ell}_S\otimes\ket{\ell}_I.
\end{align}
The fidelity is then given by
\begin{equation}
F={\rm Tr}\left(\sqrt{\sqrt{\rho}\ket{\Phi}\bra{\Phi}\sqrt{\rho}}\right).
\end{equation}
For the density matrix shown in Fig.~\ref{real}, this fidelity was found to be 83.1\%.

\begin{figure} [h]
\includegraphics[scale = 1]{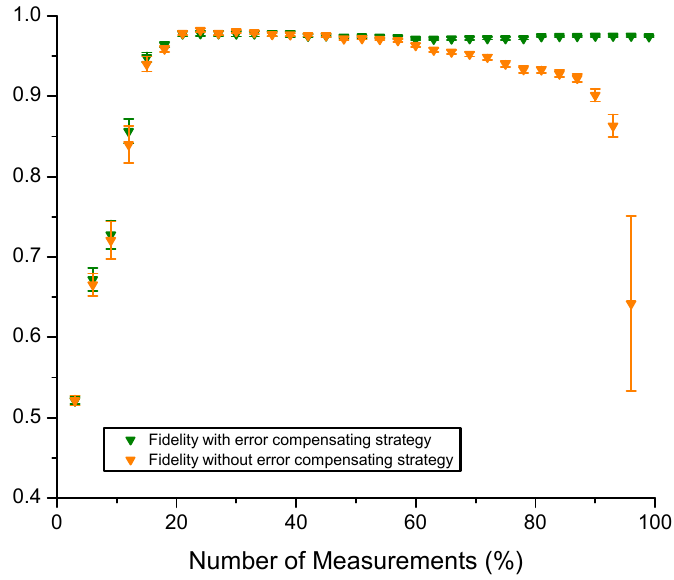}
 \caption{\footnotesize{{Fidelity with the maximally entangled state vs.~the percentage of measurements used for reconstruction for dimension $d = 7$ (100\% corresponds to $d^4$ random projective measurements).   The orange points correspond to the fidelities without error compensation; the green points correspond to the fidelities with our error compensation.  The maximum value for the green points is $0.97 \pm 0.01$.}}}
 \label{fidelity}
\end{figure}

In order to characterise the effectiveness of the reconstruction method, we reconstructed the density matrix of a 7-dimensional two-photon state with varying number of measurements. The resultant fidelities are shown in Fig.~\ref{fidelity}.  We show the results both with and without our error correction procedure.  For both cases, the fidelity increases as the number of measurements increases, indicating that more information produces a more accurate reconstruction.

However, for the case without error correction, the fidelity gradually decreases beyond 20\% of the measurements. Because the measurements performed are nearly orthogonal to each other and are of insufficient number to yield a fully determined system, the errors contained within each measurement result do not average out to reduce the uncertainty, but instead sum to increase the uncertainty. Equivalently, every measurement taken into account restricts by one dimension the space of possible solutions to the underdetermined system: fidelity increases with increasing measurements at a low number of samples because the space is large enough to be very close to the desired solution, but the space gets smaller with increasing measurements, progressively excluding other low-rank sparse objects. At the high fidelity peak, the space is small enough so that the lowest rank and sparsest solution it contains is approximately the desired one and the algorithm will converge towards it. As the number of samples increases, the accumulation of errors results in a solution space that is far from the desired one; however, with additional samples, the dimension of the space is reduced. As a result, its distance from the sampled object increases and the algorithm yields an answer that diverges from the desired one. 

\vspace{.5 cm}
{\large \noindent{\bf{Discussion}}}

\noindent
We have developed and tested an efficient method for determining the state of a quantum system based on a few simple assumptions. In this case, we use the prior knowledge of the sparsity of the density matrix associated with the system to achieve high-fidelity recovery from a small number of independent measurements of that system. Thus the state that we report corresponds to that which satisfies the set of measurements and the initial assumption of purity.  One way to look at this is to say that we have answered the following question: ``What is the purest state that is compatible with the set of measurements?" However, a feature of our method is that, using the same measurements and different prior knowledge, it can be readily refashioned to recover many states with a variety of desired characteristics.

In the case of a two-photon entangled state, where each photon exists in a 17-dimensional space with 83521 corresponding unknowns, we are able to recover the system with 3\% of the measurements required for informational completeness of an unknown general quantum state.  Our result corresponds to one of the largest discrete quantum states yet to be reported.   We anticipate that the techniques implemented in this work will have impact in a wide range of areas in quantum science, including implementation and verification of quantum information protocols using high-dimensional states.

\vspace{.5 cm}
{\large \noindent{\bf{Methods}}}

\begin{figure}[h]
\includegraphics[]{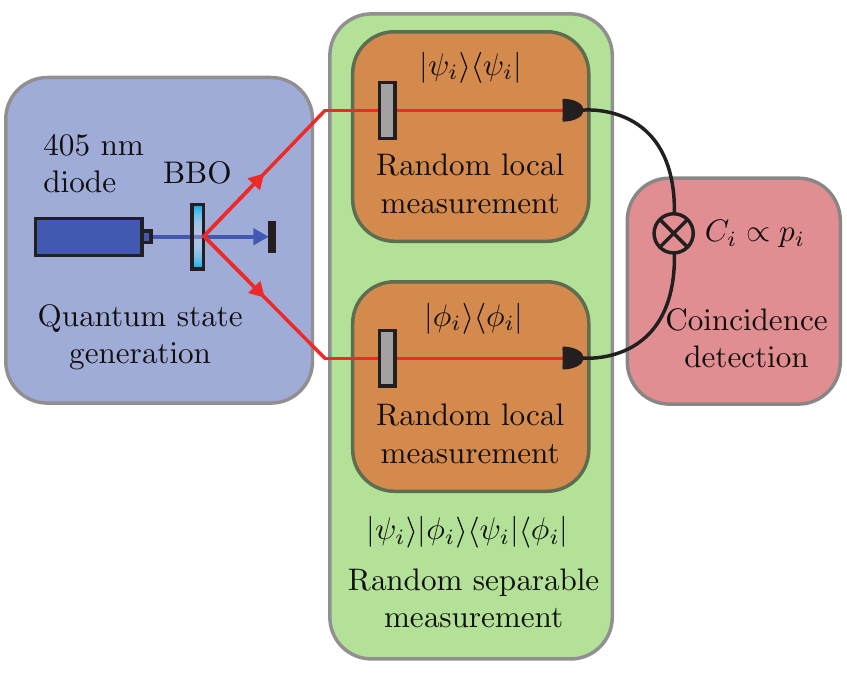}
 \caption{\footnotesize{{ Experiment configuration for compressive sensing of high-dimensional quantum states entangled in the orbital angular momentum degree of freedom.}}}
 \label{expfigure}
\end{figure}

\noindent We use a 100-mW diode laser with wavelength 405 nm, along with a 3-mm-thick BBO crystal, to generate entangled photons through the process of parametric downconversion; see Fig.~\ref{expfigure}.  The two-photon state that is generated in this process is given by 
\begin{align}
\ket{\Psi} =\sum_{\ell}c_{\ell} \ket{-\ell}_S \otimes \ket{\ell}_I,
\end{align}
where $|c_\ell|^2$ indicates the probability of finding the signal photon with OAM $-\ell\hbar$ and the idler photon with OAM $\ell\hbar$. In our experiment, we limit the range of OAM states to values between $\ell = -8$ and $\ell = 8$.

The signal and idler photons are each incident on a separate half of a spatial light modulator (SLM), displaying computer-generated holograms, and then collected by a single-mode fibre connected to a single-photon detector.    This results in a projective measurement on the two-photon mode.    The result of the projection is measured by the coincidence detection with a coincidence window of 25 ns.     

One of the keys to successful compressive sensing is to ensure that the measurement states are unstructured with respect to the basis in which the sampled state is sparse.  For this experiment, that corresponds to measurement settings that are random superpositions of OAM modes.  Therefore the measurement states $\ket{\phi_i}_S$ and $\ket{\phi_i}_I$ are generated from superpositions of OAM states where the coefficients $a_{\ell}$ are generated at random 
\begin{align}
\ket{\phi} =\sum a_{\ell} \ket{\ell}.
\end{align}

We generate the matrix $A$, of Eq.~(\ref{stcompress2}), by  performing a number of random separable projective measurements.    Each row of $A$ corresponds to the vector form of the individual projectors $\hat{A}_i$.   The projection operator $\hat{A}_i$ is given by
\begin{align}
\hat{A}_i = \ket{\phi_i}_S \ket{\phi_i}_I \bra{\phi_i}_S \bra{\phi_i}_I,
\end{align}
where the states $\ket{\phi_i}_S$ and $\ket{\phi_i}_I$ are the modes for the signal and idler arms respectively.  

The coincidence rate $c_i$ for each measurement $\hat{A}_i$ can be normalised to obtain the equivalent probability $p_i$. Each probability $p_i$ constitutes the result of the corresponding measurement $\hat{A}_i$. The probability of recording a coincidence count  is given by
\begin{align}
p_i={\rm Tr} [ \hat{A}_i \hat{\rho} ].
\end{align}
Consequently, the linear system that is defined by the set of measurement operators $\{ \hat{A_i} \}$ and the corresponding probabilities $\{ p_i  \}$ is 
\begin{align}
A \vec{\rho} = \vec{p},
\end{align}
where $\vec{\rho}$ is the vector form of the density matrix $\hat{\rho}$.  After performing an appropriate number of measurements, the task is then to solve the inverse problem under the constraints given by Eq.~(\ref{stcompress2}).


\section{Supplementary Materials}

\subsection{Algorithm Schematics}

\textbf{Input:} Matrix of measurements $A \in \mathbb{C}^{M \times N}$, vector of normalised probabilities $\vec{p}$, initial guess matrix $\hat{\rho}_{in}$, SVD threshold parameter $\tau$ such that $0 < \tau < 1$, sparsity parameter $\tau_{\ell}$ such that $0 < \tau_{\ell} < 1$ and stopping condition step size $\delta_s$. In our 17-dimensional state reconstruction we choose $\tau = 0.4$, $\tau_{\ell} = 0.04$ and $\delta_s = |\vec{\rho}_s| \cdot 10^{-3}$. 

\textbf{Output:} Recovered matrix $\hat{\rho}_d$

\begin{enumerate}
  
  \item Set $A' = orth(A)$ and $\vec{p}^{\,\prime} = C \vec{p}$ where $C A = A'$
  \item Set $\hat{\rho}_{s,0} = \hat{\rho}_{in}$
  \item \textbf{For} $k=1:k_{max}$
  \item \quad \quad Set $\hat{\rho}_{0,k} = \Gamma_{\tau, \tau_{\ell}}(\hat{\rho}_{s,k-1})$ 
  \item \quad \quad Set $\vec{\rho}_{0,k} =vec( \hat{\rho}_{0,k})$
  \item \quad \quad \textbf{For} i=1:M
  \item \quad \quad \quad \quad Set $A_i = $ $i_{th}$ row of $A$
  \item \quad \quad \quad \quad Set $\vec{\rho}_{i} = P(\vec{\rho}_{i-1},A_i)$ 
  \item \quad \quad \textbf{End}
  \item \quad \quad Set $\hat{\rho}_{s,k} = mat(\vec{\rho}_{M})$
  \item \quad \quad Set $\delta = | \hat{\rho}_{s,k} - \hat{\rho}_{s,k-1} |$
  \item \quad \quad \textbf{If} $\delta \leq \delta_s$ \textbf{Break}
  \item \textbf{End}
  
\end{enumerate}

Here $orth(\cdot)$ is the orthogonalizing operator, $\Gamma_{\tau, \tau_{\ell}}( \cdot )$ is the operator that enforces the desired characteristics described in the results section, $vec(\cdot)$ is the operator that rearranges the elements of a matrix into a vector, $P(v,V)$ denotes the projection of a vector $v$ onto a hyperplane having normal $V$, and $mat(\cdot)$ is the operator that rearranges the elements of a vector into a square matrix.

\subsection{Projection onto a hyperplane}

To project a point $\vec{\rho}_{i-1}$ onto a hyperplane $A^{\prime}_{i}\vec{\rho} = p^{\prime}_{i}$, it is necessary to find the vector $\vec{v}_i$ that has direction $\vec{n}_i$ normal to the hyperplane $A^{\prime}_{i}\vec{\rho}=p^{\prime}_{i}$ and size $k_i$, where $k_i$ is

\begin{align}\label{stcompress4}
k_i = p^{\prime}_{i}-\braket{\vec{n}_i}{\vec{\rho}_{i-1}}.
\end{align}

The desired projection $\vec{\rho}_i$ is then

\begin{align}\label{stcompress5}
\vec{\rho}_i = \vec{\rho}_{i-1} + \vec{v}_i.
\end{align}

\subsection{Finding and correcting $\Delta\vec{\rho}$}

 The error vector $\Delta\vec{\rho}$ depends on the experimental error $\Delta p_i$ associated with each of the probabilities and the measurements made to perform the reconstruction. Instead of extending the search to a non-linear space, we make use of the low rank and sparsity  information to estimate the error direction, that is, to find a close approximation for the vector $\Delta\vec{\rho}=\vec{\rho}_r - \vec{\rho}_d$, where $\vec{\rho}_r$ is the projection of $\vec{\rho}_d$ onto the linear space intersection of all the hyperplanes. In order to do this we divide the measurements and the corresponding probabilities in subsets $A_{s_i}$ and $\vec{p}_{s_i}$ of sufficient size for our compressive sensing technique to yield convergence to a solution (the size of the subsets depends on the purity of the state and the estimate of the error $\Delta p_i$ on each of the probabilities). We then perform the operation-projection algorithm on each subset separately to find the vectors $\vec{\rho}_{s_i}$, low rank and sparse solutions to the corresponding subsets systems $A_{s_i}\vec{\rho} = \vec{p}_{s_i}$. We hence define a new set of vectors $\Delta \vec{\rho}_i = \vec{\rho}_{s_i} - \vec{\rho}_{d_i}$ where $\vec{\rho}_{d_i}$ is the solution resulting from applying one last set of thresholding operations to $\vec{\rho}_{s_i}$. The sum of the vectors $\Delta \vec{\rho}_i$ is taken as a close approximation for the error vector $\Delta\vec{ \rho}$. We finally compute a correction vector for the probabilities $\Delta p$ that can be subtracted from the measured probabilities $p$. The correction vector is defined as
 
 \begin{align}\label{stcompress6}
\Delta p = A \Delta \vec{\rho}.
\end{align}

Once the probabilities have been corrected as described above, the compressive sensing algorithm can be performed, making use of the set of performed measurements $A$ and the corrected probabilities $p-\Delta p $.

\end{document}